\documentclass[prd,preprintnumbers,floatfix,aps,nofootinbib,notitlepage]{revtex4}
\usepackage{color}
\usepackage{latexsym}
\usepackage{epsfig}
\usepackage{amssymb}

\newcommand{\with}{\mx{ with }}

\newcommand{\LF}{\left(}
\newcommand{\RF}{\right)}
\newcommand{\LT}{\left[}
\newcommand{\RT}{\right]}

\newcommand{\cO}{{\cal O}}

\newcommand{\cP}{{\cal P}}
\newcommand{\cR}{{\cal R}}

\newcommand{\lp}{\left(}
\newcommand{\rp}{\right)}
\newcommand{\lb}{\left[}
\newcommand{\rb}{\right]}

\newcommand{\ba}{\begin{eqnarray}}
\newcommand{\ea}{\end{eqnarray}}
\newcommand{\be}{\begin{equation}}
\newcommand{\ee}{\end{equation}}
\newcommand{\bk}{{\bf k}}

\newcommand{\br}{{\bf r}}
\newcommand{\bx}{{\bf x}}
\newcommand{\bn}{{\bf n}}

\newcommand{\lx}{\langle}
\newcommand{\rx}{\rangle}

\newcommand{\Ve}{\e{V}}
\newcommand{\e}{\overline}
\newcommand{\ie}{{\it i.e.\ }}
\newcommand{\Le}{\e{L}}
\newcommand{\kb}{{\bf k}}

\newcommand{\bea}{\begin{eqnarray}}
\newcommand{\eea}{\end{eqnarray}}
\newcommand{\dw}{\w{\da}}

\newcommand{\w}{\widetilde}
\newcommand{\where}{\mx{ where }}
\newcommand{\xb}{{\bf x}}
\newcommand{\rr}{{\bf r}}

\newcommand{\2}{\frac{1}{2}}
\newcommand{\p}{\partial}

\newcommand{\bt}{\beta}
\newcommand{\ga}{\gamma}

\newcommand{\da}{\delta}

\newcommand{\za}{\zeta}

\newcommand{\Da}{\Delta}
\newcommand{\Oa}{\Omega}

\newcommand{\h}{\mathcal{H}}

\newcommand{\cof}{A}
\newcommand{\mx}{\mbox}
\newcommand{\mt}{\mathtt}
\newcommand{\mand}{\mx{ and }}
\newcommand{\non}{\nonumber\\}
\begin{document}

\title{Cosmological perturbations from statistical thermal fluctuations}

\author{Tirthabir Biswas}
\email{tbiswas@loyno.edu}
\affiliation{Department of Physics, Loyola University, New Orleans, LA 56302}
\author{Robert Brandenberger}
\email{rhb@hep.physics.mcgill.ca}
\affiliation{Department of Physics, McGill University, Montreal, QC H3A 2T8, Canada}
\author{Tomi Koivisto}
\email{tomik@astro.uio.no}
\affiliation{Institute of Theoretical Astrophysics, University of Oslo, P.O.\ Box 1029 Blindern, N-0315 Oslo, Norway}
\author{Anupam Mazumdar}
\email{a.mazumdar@lancaster.ac.uk}
\affiliation{Consortium for Fundamental Physics, Physics Department, Lancaster University, LA1 4YB, UK}

\date{\today}

\begin{abstract}
Cosmological perturbations due to statistical thermal fluctuations in a single fluid characterized by an arbitrary equation of state are computed. Formulas to predict the scalar and tensor perturbation spectra and nongaussianity parameters at a given temperature are derived. These results are relevant to any cosmological scenario where cosmic structures may have been seeded thermally instead of originating purely from quantum vacuum fluctuations.
\end{abstract}

\maketitle

\section{Introduction}

In the standard inflationary paradigm, the seeds for cosmic structure are generated as quantum fluctuations. During inflation, the quantum fluctuations of the fields present are stretched by the cosmic expansion to macroscopic sizes and become
classical~\cite{Brandenberger}. These small inhomogeneities are then amplified in the later evolution of the universe by gravitational collapse and eventually form the galaxies and other structures we observe around us today. The predictions of simplest inflationary models can be matched with observations that require a nearly scale invariant but slightly red-tilted spectrum with {\it only} upper limits having been set on gravitational waves and any deviations from the simplest statistical properties in terms of nongaussianity or statistical anisotropy~\cite{WMAP}.

Thermal fluctuations introduce another possible origin for small inhomogeneities and anisotropies. Thermal fluctuations are different from fluid hydrodynamical fluctuations~\cite{Brandenberger,Garriga}. In general, fluid fluctuations can arise from two different sources: There can be fluctuations in energy density and the associated temperature driven, for instance, by quantum fluctuations; this is what is traditionally discussed in literature. However, even if one can define a unique temperature in a given volume, there are fluctuations in energy within the volume due to the statistical nature of thermal physics. These are random fluctuations in all finite temperature systems that arise already at the classical level, and this is what is commonly referred to as thermal fluctuations~\footnote{In appendix~\ref{sec:comparison}, we provide the condition when the statistical thermal fluctuations dominates over the fluid fluctuations, clarifying some of the physics issues in the process.}. In the early universe the temperatures could be very high, and therefore these fluctuations could be significant. The reason  why in typical inflationary scenarios we do not worry about these fluctuations is that once inflation begins any ``pre-inflationary'' thermal matter is expected to dilute away rapidly leaving us with an almost pure vacuum state. However, there are  cosmological models where thermal fluctuations could be solely or to a significant amount responsible for the initial seeds of inhomogeneities. 
For instance, in cyclic inflationary scenarios ~\cite{cyclic, cyclic-prediction}, where particle/entropy production keeps up with the inflationary dilution, thermal fluctuations become relevant~\footnote{For other inflationary scenarios where thermal matter is relevant see \cite{BasteroGil:2009ec,warm}.}, for some of the interesting results we have found the reader is referred to our companion paper ~\cite{companion}. 
In bouncing cosmologies, where the big bang singularity is replaced by a smooth evolution from a contracting to an expanding phase, different matter sources become important near the bounce, for a recent review see~\cite{matter-bounce}, again making the thermal fluctuations relevant.

In the present study we shall not refer to particular models but rather strive for generality. The purpose is to develop the general formalism to tackle cosmological perturbations due to thermal fluctuations. We are going to make the following assumptions:

\begin{itemize}

\item The universe contains a thermal fluid in ``significant'' abundance, \ie the interactions within the fluid are able to maintain thermal equilibrium, this requires both {\it kinetic} and {\it chemical equilibrium}, for discussions, see~\cite{infl-rev,preheating-rev}. Typically this means that the relevant scattering rates have to be larger than the Hubble expansion rate. This requirement is often referred to as the ability of the fluid to maintain ``local'' thermal equilibrium.

\item For the sub-Hubble modes, (\ie physical wavelength smaller than the Hubble radius, or the appropriate cosmological time-scale), the statistical thermal fluctuations dominate over the quantum vacuum fluctuations inherent in the fluid. We will provide a quantitative criteria for this to occur in appendix~\ref{sec:comparison}.

\item There is no significant isocurvature perturbations due to the possible presence of other fluids.

\item There are no anisotropic stresses in any of the fluids.

\item There is some cosmological mechanism in place for the modes to exit from the sub-Hubble to the super-Hubble phase, after which the fluctuations evolve according to the usual hydrodynamical equations coupled to gravity. For instance, this could be attained by inflation where the Hubble radius stays approximately constant while the modes grow quasi-exponentially, or  during a regular contraction phase where the Hubble radius shrinks faster than the contraction of the modes, as in the ekpyrotic~\cite{Khoury:2001wf} or matter bounce~\cite{matter-bounce} scenario. For some other ways of realizing such a mechanism see~\cite{kinney}. As previous literature dealing with thermal fluctuations~\cite{pogosian,nayeri,param,Cai-thermal}, we also assume that the transition from sub- to super-Hubble phase is instantaneous. To study the precise nature of the transition would involve non-equilibrium thermodynamics in curved space-time which is clearly out of the scope of the present paper, but we do not expect the results to be affected beyond $\cO(1)$ factors.

\item At least, near the sub to super transition we can trust General Relativity and the usual laws of thermodynamics.
\end{itemize}

In previous literature thermal statistical fluctuations have been considered in a  variety of contexts: The earliest hint that these could be relevant for CMB was perhaps provided by Peebles~\cite{peebles}, and further developed in~\cite{pogosian,Hall:2003zp,param} (see also \cite{Taylor:2000ze}). Since then thermal fluctuations have found applications in several models: string cosmology~\cite{nayeri,BBMS}, inflation inflation (in particular warm inflation \cite{Berera:1995wh,Berera:1995ie} - for reviews see e.g. \cite{BasteroGil:2009ec,warm} - but see also ~\cite{freese,Chen,Lieu,Jinno}), bouncing cosmologies~\cite{Cai-thermal} Milne/holographic universe~\cite{Magueijo-milne,Magueijo-holography,Wu-thermal}. In the present paper we first generalize the calculation of the curvature perturbations to the case when the thermal matter could have an arbitrary equation of state~\footnote{This is going to be particularly relevant for applications to cyclic inflation models.}, in the process clarifying several conceptual issues related to gauge choices, and transfer of perturbations from sub- to super-Hubble phase. These generalizations can be particularly important and interesting for early universe cosmology where stringy
 thermodynamics~\cite{Atick-witten,Deo,Vafa-Tseytlin,Vafa-robert,BCK,BCK-prl,Bluhm,BKR} and/or phase transitions may be relevant, neither of which is described by a constant equation of state parameter which is what previous studies have been mostly confined to.  Next, we provide general formulas to compute not only the scalar power spectrum, but also the spectrum of gravity waves and higher point correlation functions. These results when applied to phase transitions in cyclic inflationary models turn out to produce interesting signatures for Planck, and will be discussed in~\cite{companion}.

The paper is organized as follows. In section \ref{sec:curv} we will derive the curvature perturbation in a universe filled with a thermal fluid with an arbitrary equation of state. We also compute the spectrum of gravity waves expected in this set-up, or equivalently the tensor-to-scalar ratio. These derivations are somewhat technical, but only familiarity with standard cosmological perturbation theory is assumed\footnote{For a very pedagogical and transparent introduction, see theory.physics.helsinki.fi/$\sim$genrel/CosPerShort.pdf}. In section \ref{sec:NG}, we will derive the nongaussianity parameters due to the thermal fluctuations, and in section~\ref{sec:radiation} we illustrate the application of our formulas for a radiation dominated contracting universe. Section \ref{sec:conc} briefly concludes. Appendix \ref{sec:window} concerns a technical issue of going over from real to Fourier space that is needed to make contact with the usual cosmological perturbation analysis; in appendix~\ref{sec:comparison} we compare the relative strengths between quantum/hydrodynamical and statistical thermal fluctuations, and in appendix~\ref{sec:pressure} we calculate  thermal pressure fluctuations for completeness.
\section{The curvature perturbation from thermal fluctuations}
\label{sec:curv}

\subsection{Curvature Perturbation \& the appropriate Gauge Choice}
\label{sec:zeta-rho}

We are going to consider a cosmological set-up where the dominant fluid component of the universe is thermal, i.e. there exists local thermal equilibrium, so that as long as the wavelengths of fluctuations are smaller than the cosmological time scale their power spectrum is determined by the thermal fluctuations in the thermal fluid.  Once the modes become super-Hubble, thermal correlations over the relevant physical wavelengths can no longer be maintained, instead the fluctuations evolve according to the usual hydrodynamical differential equations coupling the metric and the matter fluctuations. Essentially, in this set-up the thermal fluctuations act as initial conditions to seed the super-Hubble fluctuations.

Now, the super-Hubble modes are easy to track because they behave as zero
modes and it is well known~\cite{Brandenberger} that the curvature perturbation, $\zeta_k$, remains a
constant even if the equation of state parameter is not~\footnote{In
a contracting universe one of the two modes of $\zeta_k$ is growing (the
one which is decaying in an expanding universe), while the second is
constant~\cite{wands, FB}. We will here assume that the growing
mode in the contracting phase couples only to the decaying mode in
the expanding phase. Whether this is the case or not depends on
the specific model of the transition between contraction and
expansion. For a discussion of this issue see e.g.~\cite{Vernizzi}.}.
In fact, the above statement is true even if General Relativity is not
valid~\cite{Cardoso}, but as long as we are only looking at adiabatic super-Hubble
perturbations. This makes our analysis applicable to several
bouncing/cyclic models which resort to modifying gravity to obtain a
non-singular bounce (modulo the caveats about mode mixing mentioned
in the previous footnote). However, if the universe does contain more than
one type of fluid/fields, then isocurvature perturbations could be important,
but we are not going to consider them in this paper.

Our goal in this section will be to compute $\zeta_{k}$ arising from thermal
fluctuations in the sub-Hubble phase where thermal correlations can exist.
In particular we will evaluate this at the Hubble crossing which, according to our
previous discussion, will provide us with the primordial spectrum for CMBR. In
the next section, we are going to
calculate the two and the three point correlation functions as well as the
gravity wave spectrum.
We do not make any assumptions about whether we have an expanding or a
contracting universe (again modulo the previous comments about mode
matching) or whether there
is a single thermal fluid or several energy components. However, we shall make
the crucial assumption that the
fluctuations are dominated by the thermal fluctuations of a single fluid, $\delta \rho$, where $\rho$ is the energy density of the thermal fluid. We parameterize the contribution of the thermal fluid to the energy budget  by
\be \label{omega}
\Omega = \frac{a^2\rho}{3M_p^2\h^2}\,,
\ee
where $a$ is the scale factor of the universe, $\h =\dot{a}/a$ is the conformal Hubble rate, $M_p$ is the reduced Planck mass, and here and in the following the overdot denotes the derivative with respect to conformal time. Thus, if the thermal fluid is the only component in the universe, $\Omega=1$.

We would now like to relate the perturbations in the fluid to a gauge invariant degree of freedom describing the metric perturbations.
The scalar perturbations in the metric can be parameterised as~\cite{Brandenberger}
\be
ds^2 = a^2(\eta)\lb -(1+2\phi)d\eta^2 + B_{,i}d\eta dx^i + \lp (1-2\psi)\delta_{ij} + E_{,ij} \rp dx^i dx^j\rb \,.
\ee
We need also to parameterize the perturbations in the matter content, which will be treated as a perfect fluid, so no anisotropic stresses are present. Then the energy momentum tensor can be written as
\be
T^0_{\phantom{0}0} = -\rho\lp 1 + \delta\rp\,, \quad
T^0_{\phantom{0}i} =   \rho (1 + w)v_{,i} \,, \quad
T^i_{\phantom{i}j} =   \rho \lp w + c_s^2\delta\rp \,.
\ee
Since we assume a thermal fluid, all the background quantities are given solely by the temperature,
\be
\rho=\rho(T)\,, \quad p=p(T)\,.
\ee
and so is $w(T)$.
and $c_s^2(T)$~\footnote{Unlike in the case of hydrodynamical fluctuations, the pressure fluctuation is not given by the adiabatic value, $\da p = (p'(T)/\rho'(T))\da \rho$, but is rather determined via stress-energy conservation, see
appendix~\ref{sec:pressure} for the precise formula. We will see shortly, that to compute the spectra we do not need the explicit form of the sound speed squared. However, as the modes exit the horizon, the perturbations are expected to relax to their adiabatic value which guarantees the constancy of the curvature perturbation at large scales.}.
In the following the prime will always denote the derivative with respect to the temperature, and we often drop the explicit argument $T$. An overdot will refer to a derivative with respect to the conformal time $\eta$.

Now we want to calculate the gauge invariant curvature perturbation, which  in a
general gauge reads as
\be \label{zetad}
\zeta = -\Psi - \h (v-B)\,.
\ee
Since it is a gauge invariant quantity we can evaluate it in any gauge of our choice. Let us choose to work in the longitudinal gauge where $E=B=0$. In this gauge the metric can be written in terms of the gauge invariant Bardeen potentials $\Phi$ and $\Psi$:
\be
ds^2 = a^2(\eta)\lb -(1+2\Phi)d\eta^2 + (1-2\Psi)dx^2\rb \,,
\ee
and  the $0i$ component of the Einstein's equations determines $v$ in terms of the Bardeen potentials:
\be
\dot{\Psi}+\h \Phi = \frac{a^2}{2M_p^2}(1+w)\rho v\,.
\ee
Using this in (\ref{zetad}) we have that
\be
\zeta = -\Psi - \frac{2M_p^2\h}{(1+w)a^2\rho}\lp \dot{\Psi} + \h\Phi\rp\,.
\label{zeta-bardeen}
\ee

Please note that the above equation is written in a completely gauge invariant form and is therefore valid in any gauge. This is important for us because it will become necessary for us to switch to the comoving gauge on physical grounds. This is actually a subtle issue which, to our knowledge, has not been explained before. The point is that all thermodynamic calculations, such as those relevant when we will derive the energy fluctuations in a given volume,  are typically carried out in Minkowski space-time. In order to generalize the analysis to the FLRW metric (or any other metric for that matter) one has go to a frame where the background fluid is at ``rest''~\cite{misner-thorne-wheeler}, which is none other than the comoving gauge. In this gauge $\da \rho$ becomes gauge-invariant, as it must because the Minkowski calculations cannot obviously depend on the gauge choice, and is related to the Bardeen potential via the relativistic Poisson equation:
\be
\Psi = - \frac{1}{2}\lp \frac{a}{kM_p}\rp^2\delta\rho^C\,.
\label{poisson}
\ee
The superscript ``$C$" refers the comoving gauge which we are going to subsequently drop as all the thermodynamic calculations implicitly assumes this same gauge choice for the perturbation in the matter density field.

At this point it is useful to set $\Phi=\Psi$, since we assumed that the
anisotropic stresses can be neglected. We can then compute $\dot{\Phi}_k$ in terms of the density fluctuations from (\ref{poisson}) and substitute it in  (\ref{zeta-bardeen}) to obtain
\be
\zeta =\frac{1}{2}\lp \frac{a}{kM_p}\rp^2\lb 1+\frac{2M_p^2H^2}{(1+w)\rho}(3+r)\rb\delta\rho\,,
\ee
where time evolution of the density fluctuation is parameterized as
\be \label{rate}
r = \frac{d \log{\delta\rho}}{d\log{a}} = \frac{(\delta\rho)'}{\delta\rho} \frac{\dot{T}}{\h}\,.
\ee
We remind the readers that the prime corresponds to the derivative with respect to the temperature. 

More succinctly,
\be \label{zeta-rho}
\zeta_k  =  \frac{\cof(T_k)}{H_k^2 M_p^2}\delta\rho_k\,,
\ee
where we have defined a time/temperature dependent proportionality coefficient
\be \label{zeta}
A(T) \equiv \frac{1}{2}\lb 1+\frac{2(3+r)}{3(1+w)\Omega}\rb \ ,
\ee
for later convenience. These quantities will depend on the temperature at the time of the ``exit'' of a given comoving mode.
\subsection{Thermal Density Fluctuations}
\label{sec:T-flucts}
We will now use the thermodynamics to quantify fluctuations in the fluid and then use the results of the previous section to relate them to the metric perturbation spectra. This is more of a review of what has been discussed in the previous literature~\cite{nayeri,param,Cai-thermal}, and our results agree to within $\cO(1)$ factors, until a crucial step highlighted in the end this subsection. 

One defines the average fluctuation in energy, $\Da E$ via
\bea
\lx \Da E\rx_L^2\equiv \lx E^2\rx  -\lx E\rx\rx  ^2&=&{1\over Z}{\p^2 Z\over \p \bt^2}-\lp {1\over Z}{\p Z\over \p \bt}\rp^2={\p^2 \ln Z\over \p \bt^2}=-{\p \lx E\rx  \over \p \bt}=T^2 C_L\ ,\\
\lx \da \rho^2\rx  _L&=& {T^2 C_V\over L^6}= {T^2\over L^{3}}{ \p \rho\over   \p T}\,,
\eea
where $C_L$ is the heat capacity of the thermal system for a given volume $L^3$. Note we have also introduced a subscript $L$ in $(\Da E)_L^2$ to denote that we are considering fluctuations in a given volume.

The next step is to go from real space to momentum space. This is a  tricky procedure as it depends to some extent on the window function one chooses. In appendix~\ref{sec:window}, we consider in details this procedure using gaussian window function~\footnote{Different schemes typically yield slightly different values for  $\gamma$.} and obtain
\be \label{spectrak}
\delta\rho^2_k = \frac{\gamma^2}{k^3}\langle \delta\rho^2 \rangle_{L=a/k} \with \ga =2\sqrt{2}\pi^{3/4}\approx 6.7\ .
\ee
 Thus we have
\be
\delta\rho^2_k =\frac{\gamma^2}{a^3}T^2\rho'
\label{del-rho}
\ee
leading to
\be \label{zeta}
\zeta_k^2 = \cof^2(T_k)\frac{\gamma^2}{a^3} \frac{T^2\rho'}{H_k^4 M_p^4}\,,
\ee
and eventually
\ba
\cP_{\za}=k^3\langle \zeta_k^2 \rangle= \cof^2(T_k)\gamma^2 \frac{T_k^2\rho_k'}{H_k M_p^4} =  \sqrt{3\Omega}\gamma^2\cof^2(T_k)\frac{T_k^2\rho_k'}{M_p^3\sqrt{\rho_k}}\,. \label{spectrum}\ ,
\ea
using the standard definitions of the power spectrum. The subscript $k$ refers to the fact that all these quantities have to be evaluated at the Hubble crossing condition, $H_k=k/a$, which we have also used along with (\ref{omega}) to eliminate the Hubble factors. A few comments are now in order. Firstly, as in all previous literature on the subject in deriving (\ref{spectrum}) we have implicitly assumed an instantaneous transition from the thermally correlated sub-Hubble phase to the hydrodynamical super-Hubble phase. This is obviously not realistic but a careful investigation of such a transition is very challenging  because it will involve non-equilibrium thermodynamics on curved space-time, clearly out of the scope of the present paper. 

Secondly, one can see that the factor $\cof(T)$ basically tells us what is the difference between the spectra of the gravitational potential (or, via the Poisson equation, the density perturbation) and the spectra of the gauge invariant curvature perturbation. It is crucial to take this into account: one can imagine physical situations where $\cof$ can even vanish or diverge. This is where our results differ significantly from previous literature which typically only computes $\Phi_k$ and are directly applicable only for constant equation of state parameters when making comparison with observations.

\subsection{The prefactor $\cof(T)$}

The last missing piece required to obtain the power spectrum is an expression for $A(T)$. Explicitly, our definition is
\be
\cof(T)=\frac{3(1+w)\Omega + 2(3+r)}{6(1+w)\Omega}\,.
\ee
$w$ can be obtained straight forwardly as a function of temperature from the partition function, or $p(T)$. A useful thermodynamic relation, in this context is
\be
\rho(T)=T{d p(T)\over d T}-p(T)
\ee
so that
\be
w={p\over T p'-p}
\ee
The computation of $\Oa$ at the exit temperature depends on the specific model under consideration, and one cannot make any further simplifications at this point. Obviously, if the thermal fluid is the only energy component in the universe, then $\Oa=1$.

We are finally left with the evaluation of $r$.  From the expression we obtained for thermal energy density fluctuations (\ref{del-rho}) we first find
$$
r=-{3\over 2}+\LF{2\rho'+T\rho''\over 2\rho'}\RF{d\ln T\over d\ln a}
$$
Now, recalling the continuity equation
\be
\dot{\rho}+3\h\lp 1 + w\rp\rho = 0\,,
\ee
we see
\be \label{cont}
\frac{d\ln T}{d\ln a}=-3(1+w)\frac{\rho}{T\rho'}\,.
\ee
Thus we finally have
\be \label{rate2}
r=-{3\over 2}\LT1+\frac{\lp1+w\rp\rho\lp2\rho'+T\rho''\rp}{T{\rho'}^2}\,\RT.
\ee
Note that the sign of this remains the same also in the contracting phase: then the temperature is getting lower with time, but also the
Hubble rate is negative. We will illustrate the computation of the power-spectrum for the special case of radiation towards the end of the next section.


\subsection{Gravity Waves}
Another interesting probe of our early universe is the primordial gravitational wave which are stretched like the scalar perturbations.
However, for a linearized Einstein's gravity there is no source term for the gravitational waves. In principle the initial conditions for the gravitational waves could be set purely classically~\cite{amjad} or from quantum vacuum condition~\cite{Brandenberger}~\footnote{In fact, any thermal matter can only act as sources of gravitational waves if it's partition function is non-extensive~\cite{Deo,nayeri}, a scenario not considered here.}.  Assuming that the initial conditions for the primordial gravitational waves are set by quantum vacuum, i.e. Bunch-Davis, the gravitational wave spectrum is given by
\be
\mathcal{P}_h = \frac{1}{4\pi^2}\lp \frac{H}{M_p}\rp^2 = \frac{\rho}{12\pi^2 M_p^4 \Omega}\,.
\ee
The tensor to scalar ratio is then given by
\be
r_{t/s}\equiv {\cP_h\over \cP_{\za}} = \frac{1}{\ga^2} \frac{1}{12\sqrt{3}\pi^3\Omega^\frac{3}{2}\cof^2(T)}\frac{\rho^\frac{3}{2}}{M_pT^2\rho'}\,. \label{rst}
\ee
In general, the temperature dependence of this and the scalar spectra too depend very nonlinearly on the properties of the thermal fluid, but these are straightforward to compute once we know $\rho(T)$.

\section{Nongaussianity: bi-and tri-spectrum for the curvature perturbations}
 \label{sec:NG}

In the previous section we evaluated the CMB power spectrum as a two step process. In section~\ref{sec:T-flucts} we calculated the thermal density fluctuations from the partition function (or pressure) governing the thermodynamics of the fluid in the comoving gauge in which the fluid is ``at rest'' and therefore the Minkowski space-time calculations can be applied~\cite{misner-thorne-wheeler}. In section~\ref{sec:zeta-rho} we found how these thermal fluctuations are related to the curvature fluctuations, which then allowed us to obtain the two-point correlation function in CMB. We can apply the same prescription to obtain higher point correlation functions, we just have to compute the appropriate higher thermodynamic cumulants (see also \cite{pogosian} for an earlier study of non-Gaussianities from thermal fluctuations)..

The third and the fourth order centered cumulants are given by
\bea
-{\p^3 \ln Z\over \p\bt^3}&=&\lx E^3\rx  -3\lx E^2\rx  \lx E\rx  +2\lx E\rx  ^3\equiv \lx \Da E^3\rx\,,  \non\\
-{\p^4 \ln Z\over \p\bt^4}&=&\lx E^4\rx  -4\lx E^3\rx  \lx E\rx  +6\lx E^2\rx  \lx E\rx  ^2-4\lx E\rx  \lx E\rx  ^3\equiv \lx \Da E^4\rx\,.  \non
\eea

From the above thermodynamics we infer that
\ba \label{spectra}
\langle \delta\rho^3 \rangle_L & = & \frac{T^3\lp 2\rho' + T\rho''\rp}{L^6}\,, \\
\langle \delta\rho^4 \rangle_L & = & \frac{2T^4\lp 3\rho' + 3T\rho'' + \rho'''\rp}{L^9}\,,
\ea
where one considers fluctuations in a box of size $L$. These formulas hold in the real space, but can be converted to momentum space using window functions as before~\footnote{The relations to the Fourier space spectra are tricky, and the $\ga$ factors for the different correlations functions could be different depending upon the window functions used, but the difference is only expected to provide $\cO(1)$ modulations.} we have
\be \label{spectrak}
\langle \delta\rho^3 \rangle = \frac{\gamma^3}{k^\frac{9}{2}}\langle \delta\rho^3 \rangle_L \,, \quad
\langle \delta\rho^4 \rangle = \frac{\gamma^4}{k^6}\langle \delta\rho^4 \rangle_L \,.
\ee

Using the standard definitions of the spectrum and nongaussianity parameters, see~\cite{Maldacena,WMAP}, we are ready to write down the results using (\ref{zeta}) in (\ref{spectrak}):
\ba
f_{NL} & = & {5\over 8}k^{-{3\over 2}}{\lx \xi_k^3\rx  \over\lx \xi_k^2\rx  ^2}=
  \frac{1}{\Omega\ga \cof(T)}\lb \frac{5\rho\lp 2\rho' + T\rho''\rp}{24T\lp \rho'\rp^2}\rb\equiv \frac{F(T)}{\Omega\ga\cof(T)} \,. \label{f_nl} \\
g_{NL} & = & \frac{25}{54}k^{-3}\frac{\lx \xi_k^4\rx  }{\lx \xi_k^2\rx  ^3} =
\frac{1}{\Omega^2\ga^2\cof^2(T)}
\lb \frac{25\rho^2\lb 3\lp \rho' + T\rho''\rp + T^2\rho'''\rb}{243T^2\lp\rho'\rp^3} \rb\equiv \frac{G(T)}{\Omega^2\ga^2\cof^2(T)}\,. \label{g_nl}
\ea
In deriving the expressions for the nongaussianity parameters we have used (\ref{zeta-rho}) to relate fluctuations in energy density to the curvature fluctuations, and also (\ref{omega}) to eliminate the Hubble factors.

Physically, the most important aspect about the thermal NG parameters is that they depend on how the pressure/density varies as a function of the temperature. Moreover, the higher the order of the correlation function, the higher the derivatives that becomes relevant. It's not hard to realize then that if the exit temperature is close to a thermal phase transition, we might be able to see it's effects in the enhancement of NG parameters. This is precisely what we observed in the context of cyclic inflation scenario \cite{companion}. For the present, we will just illustrate the various computations of the cosmological observables by considering the ordinary relativistic fluid. In particular, we will see that for pure radiation, the nongaussianities and the tensor-to-scalar ratio are too small to be observed by the Planck experiment.

\section{Example: Radiation}
 \label{sec:radiation}
For the purpose of illustration let us consider a radiation dominated contracting universe. Since in this case the Hubble radius contracts faster (as $1/t$) than the physical wavelengths (as $1/\sqrt{t}$), the latter is pushed out of the Hubble radius and the various perturbative spectra becomes imprinted at the time of the mode-exit. We can use the formulas of the previous section to  compute the different cosmological observables.

For relativistic radiation fluid we have that
\be
\rho(T)= g T^4 \mand p(T)={g \over 3}T^4\,.
\ee
where $g$ depends on the number of relativistic degrees of freedom.
It follows immediately that
\be
w =\frac{1}{3} \mand r=-4\,, \Rightarrow \cof=\frac{1}{4}\,.
\ee
In general, these parameters need not be constant, but they happen to be in this simple case, or whenever pressure is a power-law in temperature.

The amplitude of the primordial spectrum, according to our convention,  is then given by
\be \label{r_spectrum}
\cP_{\zeta}  = \frac{\sqrt{3 g}\ga^2}{4}\lp\frac{T}{M_p}\rp^3\,,
\ee
where $T$ corresponds to the temperature when the given mode exits becomes super-Hubble. Evidently, the spectrum is not scale invariant because the amplitude depends strongly on the temperature, and $T\propto 1/a$ 
giving rise to a very large blue tilt. We should point out that our claim
that (\ref{r_spectrum}) is the primordial spectrum relevant for CMBR relies crucially on the
fact that there is no mixing between the mode of $\zeta_k$ which is
growing in the contracting phase with the dominant mode in the expanding
phase, the constant mode, and in addition, on the assumption that there are
no entropy modes which become important and which could change the spectrum
of the curvature fluctuations on super-Hubble scales. If there
is unsuppressed mixing between the growing mode in the contracting phase
and the constant mode in the expanding phase (see~\cite{robert-rev} for examples
where this is the case), then the amplitude of the resulting curvature
fluctuations changes, but the spectrum remains as given by (\ref{r_spectrum}). The
reason that there is no change in the shape of the spectrum (in
contrast to the case of a matter-dominated phase of contraction where
the index of the power spectrum changes by $-2$, see \cite{wands, FB}) comes from the fact that for a radiative equation
of state the canonical fluctuation variable $v$ \cite{Sasaki,Mukh} which
is related to $\zeta$ via $\zeta \sim a^{-1} v$ has vanishing
squeezing factor and hence remains conserved. Thus, there is no
preferential growth of long wavelength modes compared to short wavelength
modes which would come from the fact that long wavelength modes spend
more time on super-Hubble scales in the contracting phase. For a
discussion of this point see~\cite{robert-rev}.

For radiation we obtain the following numbers for the nongaussianity parameters:
\ba
f_{NL}&=&{25\over 24\ga} \approx 0.16\,, \\
g_{NL}&=&{25\over 216\ga^2} \approx 0.003\,.
\ea
Not surprisingly for radiation, which is free from any internal scales, both the $f_{NL}$ and the $g_{NL}$ parameters turn out to be scale invariant. The above approximate values correspond to the $\ga$ value for the gaussian window function (see the appendix \ref{sec:window}) which is unfortunately beyond Planck's sensitivity.

Let us compute the tensor-to-scalar ratio for radiation.
We readily obtain
\be \label{r_rts}
r_{t/s}={4\sqrt{g }\over 75\sqrt{3}\pi^3\ga^2}{T \over  M_p}\,.
\ee
For a given temperature, we can fix the unknown
$g$ by matching the amplitude of perturbations (\ref{r_spectrum}) with the observed one. This then fixes the tensor-to-scalar ratio (\ref{r_rts}).
In other words, we can deduce the primordial temperature from observations:
\be
\frac{T}{M_p}=\frac{8}{75\gamma^2\pi^{\frac{3}{2}}}\sqrt{\frac{2A_0}{r_{t/s}}}=\frac{1}{75\pi^3}\sqrt{\frac{2A_0}{r_{t/s}}} > 6.1\cdot 10^{-8}\,.
\ee
In the second equality we have used the gaussian window value for $\ga$, and the lower bound is obtained from the present best-fit WMAP value for the amplitude
$A_0 = 2.4\cdot 10^{-9}$ and the bound $r_{t/s}<0.24$, which both apply at the scale $k=0.002/$MpC. The minimal allowed temperature corresponds to
a huge number of effective degrees of freedom, $g \sim 10^{22}$. For $g\sim 1$ of order unity, the observed amplitude requires
$T/M_p \sim 10^{-4}$, in which case the tensor-to-scalar ratio will be too low to be measured.

To conclude, thermal fluctuations in usual radiation cannot account for the CMBR spectrum as the spectrum is heavily tilted, not surprisingly. We also found that such fluctuations cannot produce large nongaussinities or gravity wave signals. Similar conclusions hold for simple polytropic thermal fluids, but as we will see in~\cite{companion} richer thermodynamics may indeed yield  detectable nongaussianities.

\section{Conclusions}
\label{sec:conc}

We considered statistical  thermal fluctuations as a possible source for cosmological large scale structures. We presented a robust derivation of scalar and tensor spectra in this context. We also explicitly provided the formulas for the bi- and trispectrum, and outlined the procedure which is straightforward to implement in order to obtain nongaussianity at an arbitrary order. The results were applied for the case of radiation for illustration, and they are easily applied to any other fluid, given nothing but its thermodynamic properties specified by the equation of state, or equivalently, the temperature dependence of its energy density. Fundamentally, these follow from the partition function.

Another question is whether there are realistic cosmological models in which statistical thermal fluctuations are dominant instead of the usual quantum fluctuations to seed the large scale structures? We believe the cyclic inflationary scenario presents a plausible framework where this indeed turns out to be the case, and in a companion paper we shall apply the results obtained here to study this scenario in detail and show that there are parameter regions compatible with the present observations and falsifiable predictions for both the tensor-to-scalar ratio and nongaussianity.

\acknowledgements{TB would like to thank Marco Peloso for raising some important issues about the nature and evolution of thermal fluctuations. TB's research has been supported by the LEQSF(2011-13)-RD-A21 grant from the Louisiana Board of Regents. TK is supported by the Research Council of Norway, AM is supported by the Lancaster-Manchester-Sheffield Consortium for Fundamental Physics under STFC grant ST/J000418/1. }

\appendix

 \section{Window Function}
 \label{sec:window}
Here we are going to take a closer look at how the fluctuations in energy in a given physical volume is converted to the power spectrum. To begin with we note that the thermodynamic computation of energy fluctuations is done in an ``adiabatic'' approximation scheme where we ignore the cosmological evolution. Thus $\da_L$ tells how in a given ``Eucledian'' time slice the ``relative energy'' in a given physical volume $L^3$ fluctuates. The question we are interested in asking is that for a given time slice, if we know how $\da_L$ depends on $L$, how can we compute the Fourier components of relative density fluctuations in that same time slice. To understand such a ``kinematical/statistical'' relationship we can  therefore work with physical coordinates and momenta. Once we derive the relationship, it is relatively easy to convert the results in ``comoving'' language which is more useful for cosmology.

To keep things finite we first choose a ``fiducial'' volume in our universe, $\Ve$, which is big enough that it contains all the relevant scales we are interested in \ie, $L\ll \Le$. We are not going to provide all the rigorous details/justifications for doing this, which are presented in~\cite{Coles}. We are also going to assume periodic boundary conditions, so that physical momenta's are confined to integral values:
\be
\kb ={2\pi\over \Le}\bn\,.
\ee
We can now define the Fourier components via
\bea
\dw_{n}&=&{1\over \Ve}\int_{\Ve} \da(x)\exp(i2\pi \bn\cdot \bx/\Le)\ d^3x\\
\da(x)&=&\sum\dw_n\exp(-i2\pi \bn\cdot \bx/\Le)\,,
\eea
and the relative density fluctuations are defined in the usual way
\be
\da(x)\equiv {\rho(x)-\lx \rho\rx  \over \lx \rho\rx  }\where \lx \rho\rx  \equiv {\int_{\Ve}\rho(x)\ d^3x\over \Ve}\,.
\ee

Now, what we are interested is to obtain a statistical measure of $\da$. However, statistically one expects
\be
\lx \da\rx  =0\,.
\ee
Thus, one has to obtain an estimate of $\da_\kb$ using a root mean square approach. The standard approach is to use the correlation function:
\be
\za(\rr)\equiv \lx \da(\xb)\da(\xb+\rr)\rx  _{\Ve}
\ee
$$=\sum_{\bn \bn'}\lx \dw_\bn\dw_{\bn'}\exp\lb-{i2\pi \bn \cdot (\bx+\br)\over\Le}\rb\exp\lb-{i2\pi \bn'\cdot \bx\over\Le}\rb\rx  =\sum_{\bn\bn'}\lx \dw_\bn\dw^{*}_{\bn'}\exp\lb-{i2\pi(\bn-\bn')\cdot \bx\over \Le}\rb\exp\lb-{i2\pi \bn \cdot \br\over\Le}\rb\rx  
$$
$$=\sum_{\bn}\lx |\dw_n|^2\rx  \exp\lb-{i2\pi \bn \cdot \br\over \Le}\rb
$$
Our final step is to go from the discretized momentum space to a continuum limit ($\Ve\rightarrow\infty$ limit). To achieve this, as is always done in statistical mechanics, we have to convert sums into integrals and go from the $n$-space to $\kb$-space:
$$
\sum_{\bn} \lx |\dw_\bn|^2\rx  \rightarrow\int d^3n  \lx |\dw_\bn|^2\rx  \equiv\int d^3\kb\ \da_\kb^2
$$
\be
\Rightarrow\da_\kb^2={\Ve\over (2\pi)^3}\lx |\dw_\bn|^2\rx  \ ,
\ee
since the $n$ and $k$ spaces are related  via the density of states:
\be
d^3n={\Ve\over (2\pi)^3}d^3\kb\,.
\ee

Thus the correlation function can now be expressed as
\be
\za(r)=\sum_{\bn}\lx |\dw_\bn|^2\rx  \exp\lb-{i2\pi \bn\cdot \br\over \Le}\rb={1\over (2\pi)^3}\int d^3\kb \da_\kb^2\exp(-i\kb\cdot \rr)\,.
\ee
In other words, $\da^2_\kb$ is just the fourier transform of the correlation function. Since the correlation function is a physical quantity \ie, it does not depend on the fiducial volume over which the averaging is performed, it is  also clear that $\da_\kb^2$ is also a physical quantity.

Using similar analysis one can also  statistically compute the mass/energy variance in a given physical volume. We find (the derivation is straight forward and given in section 13.3 of~\cite{Coles})
\be
\da_L^2=\sum_n \dw_\bn^2W^2(2\pi n L/\Le)\,,
\label{ktoL}
\ee
where the window function is defined as
\be
W\lb\kb L \rb\equiv \lb{1\over V}\int_{V}\exp\lb i\kb\cdot {\bf y}\rb\ d^3y\rb\,.
\label{bessel}
\ee
Again we can pass from the discrete $n$-space to the continuum momentum $k$-space to find
\be
\da_L^2={1\over (2\pi)^3}\int d^3\kb\ \da_\kb^2W^2(\kb L)={1\over 2\pi^2}\int d\kb\ \kb^2\da_\kb^2W^2(\kb L)\equiv{1\over 2\pi^2}\int d\kb\ \kb^2P(\kb)W^2(\kb L)
\ee

One can write down an analytical expression for the window function (\ref{bessel}) in terms of Bessel functions. However, to avoid some technical complications in literature often other window functions are used such as
\be
W_G(\kb L)\equiv \exp[-\2(\kb L)^2]\,.
\ee
We emphasize that in the above formulas $\kb$ and $L$ represent physical momenta and lengths respectively.
The inversion thus gives us
\be
\da_\kb^2=P(\kb)\sim {T^2 \p \rho\over \rho^2  \p T}\,.
\ee

The final step step involves going from the physical fourier components to the comoving fourier modes. The two are defined via
\bea
\Phi_{\vec{k}}^2={1\over (2\pi)^3}\int d^3\vec{x}\ \Phi^2(\vec{x})e^{i\vec{k}\cdot\vec{x}}\\
\Phi_{\kb}^2={1\over (2\pi)^3}\int d^3\xb\ \Phi^2(x)e^{i\bk\cdot\bx}
\eea
where we now denote the physical coordinates as vectors $\vec{x}=a(t)\bx$. It is clear now that
\be
\Phi_\bk^2=a^{-3}(t)\Phi_{\vec{k}}^2\,.
\ee
\section{Quantum vs. thermal fluctuations}
\label{sec:comparison}
To estimate the relative contributions to the density perturbations originating from thermal statistical vs. the quantum vacuum fluctuations, let us consider an ideal fluid with a constant equation of state parameter $0<w<1$. This corresponds to having a polynomial pressure as a function of temperature:
\be
p(T)= m^4\LF{T\over m}\RF^{1+w\over w} \mand p=w\rho
\ee
where $m$ is a mass scale associated with the fluid. Thus according to our formula for the spectrum, it's parametric dependence on the temperature is given by
\be
\cP_{\mt{th}}\propto \LF{m\over M_p}\RF^3 \LF{T\over m}\RF^{{3w+1\over 2w}}
\ee

In contrast, the quantum vacuum fluctuations yield a power-spectrum that is proportional to the density of the background fluid~\cite{Brandenberger}:
\be
\cP_{\mt{vac}}\propto \LF{\rho\over M_p^4}\RF\sim \LF{m\over M_p}\RF^4\LF{T\over m}\RF^{1+w\over w}
\ee
We immediately notice that the vacuum fluctuations have an extra Planck suppression:
\be
\cR\equiv {\cP_{\mt{vac}}\over \cP_{\mt{th}}}\sim \LF{T\over M_p}\RF^{{1-w\over 2w}}\LF{m\over M_p}\RF^{{3w-1\over 2w}}
\label{pert-ratio}
\ee
For instance, for radiation, since $w=1/3$, we have
\be
\cR\sim {T\over M_p}
\ee

More generally,  for the range $1\geq w\geq 1/3$, since both the exponents in (\ref{pert-ratio}) are positive, as long as $T,m<M_p$, the vacuum fluctuations are  suppressed as compared to the random thermal fluctuations. If $1/3\geq w> 0$, then depending upon the value of $w$, if $T$ is sufficiently larger than $m$, the vacuum fluctuations may be able to dominate over the thermal fluctuations. However, in most physical scenarios one expects $T\lesssim m$ for the validity of the physics involving the thermal fluid. Thus in most physical scenarios we actually expect the thermal fluctuations to dominate the show, but the formal condition for this to happen is given by
\be
\LF{m\over M_p}\RF\LF{T\over m}\RF^{{1-w\over 2w}}<1\,.
\ee

\section{Pressure perturbation}
\label{sec:pressure}
For a fluid in thermal equilibrium, the thermodynamic properties of the fluid determine the density fluctuation. This, through Einstein's field equations, is related to the gravitational potential which in turn determines the pressure fluctuation. We can thus compute the explicit form of the latter. This is a straightforward but tedious task and we omit the details of the algebra here\footnote{A convenient starting point is Eq.(15.21) in theory.physics.helsinki.fi/$\sim$genrel/CosPerShort.pdf.}. The result is
\ba
c_s^2 & = & \frac{\delta p}{\delta\rho}  =
\frac{a^2 \rho} {4 k^2 M_p^2 T^2 \left(\rho
   '\right)^4}
  \Big[9 \rho ^2 T^2 (w+1)^2 \left(\rho ''\right)^2+T^2 (5 w-2)
   \left(\rho '\right)^4 \\
   & + & \rho  (w+1) \left(\rho '\right)^2 \left(T^2 (1-15 w) \rho
   ''+24 \rho  (w+1)\right)-2 \rho  T (w+1) (15 w+2) \left(\rho '\right)^3-6 \rho ^2 T
   (w+1)^2 \rho ' \left(\rho ^{(3)} T-2 \rho ''\right)\Big]\,.\nonumber   \ea
   For radiation this yields $c_s^2=(2/3)(aT^2/kM_p)^2$.
This is certainly different from $c_{adi}^2$, which again highlights the difference between thermal and hydrodynamic perturbations. In particular, the properties of the fluid and the background expansion need to be taken into account (both of which in principle is determined by the temperature), but in addition the relation is scale dependent. In fact, approaching the small-volume limit $k\rightarrow \infty$ the pressure fluctuation becomes negligible, and at the largest-scale limit $k\rightarrow 0$ the result formally diverges, but of course thermal correlations are only expected to exist as long as $k/a<H$.

\bibliography{thermalrefs}

\end{document}